\documentclass[12pt]{iopart}

\newcommand{\RV}{R_\mathrm{V0}}
\newcommand{\SV}{\sigma_\mathrm{V0}}

\usepackage{graphicx}

\begin{document}

\title[] {Cross section normalization in proton-proton collisions at $\sqrt{s}$ = 2.76 TeV and 7 TeV, with ALICE at LHC}

\author{Ken Oyama for the ALICE Collaboration}

\address{Physikalisches Institut, University of Heidelberg, Philosophenweg 12, 69120 Heidelberg, Germany}
\ead{oyama@physi.uni-heidelberg.de}
\begin{abstract}

Measurements of the cross sections of the reference processes
seen by the ALICE trigger system were obtained based on beam
properties measured from van der Meer scans. The measurements
are essential for absolute cross section determinations of physics
processes.

The paper focuses on instrumental and technical aspects of detectors and accelerators,
including a description of the extraction of beam properties from the van der Meer scan.
As a result, cross sections of reference processes seen by the ALICE trigger system
are given for proton-proton collisions at two energies; $\sqrt{s}$=2.76 TeV and 7 TeV,
together with systematic uncertainties originating from beam intensity measurements
and other detector effects.
Consistency checks were performed by comparing to data from other experiments in LHC.

\end{abstract}

\maketitle

\section{Introduction:}

The determination of the cross section of a reference trigger process ($\sigma_{trig}$)
allows a scale normalization for other cross-section measurements in the experiment,
and enables the calculation of the luminosity $L$ using the trigger process' rate $R_{trig}$,
via a relation $R_{trig}(t) = L(t) \cdot \sigma_{trig}$.
In the ALICE experiment\cite{ALICE}, such a reference cross-section has been measured using the van der Meer (vdM) scan
method \cite{VDM}.
$R_{trig}$ is measured as a function of
the beams separation, providing information on the spatial convolution of the two colliding beams.

In heavy-ion collision experiments particle production in AA collisions is often compared
with the extrapolation from elementary $pp$ collisions via binary scaling.
The nuclear modification factor $R_{AA}^{(X)}$ for a given process $X$ is defined as the ratio between the process yield
in AA collisions $N_{AA}^{(X)}/N_{evt}$ and the yield expected by scaling the $pp$ cross-section $\sigma_{pp}^{(X)}$ by
the average nuclear overlap function $\langle T_{AA} \rangle$:
$  R_{AA}^{(X)} = (N_{AA}^{(X)}/N_{evt}) / (\langle T_{AA} \rangle \cdot \sigma_{pp}^{(X)}) $.
The desired precision for $R_{AA}^{(X)}$ studies, to quantify the importance of nuclear effects, is typically $<10$\%.
Thus a precision of the order of 5\% on $\sigma_{pp}^{(X)}$ is desired in order not to be dominant in the overall uncertainty,

\section{Detector Setup}

For the present study, the cross section ($\SV$) of the process triggered by two scintillator arrays (V0)\cite{ALICE}
has been measured as the reference.
The arrays are asymmetrically placed at $2.8 <\eta < 5.1$ (V0A) and $-1.7 > \eta > -3.7$ (V0C).
In each array the scintillator tiles are arranged in 2 (radial) $\times$ 16 (azimuthal) segments with
individual photomultiplier-tube readout.
32 logic signals of each array after applying pulse height thresholds are combined into a logical ``OR''.
The two resulting signals are combined with an ``AND'' logic with 25 ns of coincidence window,
to reduce the sensitivity to backgrounds.

\section{van der Meer Scan Analysis}

In the vdM scan\cite{VDM}, the luminosity $L$ (thus trigger rate $\RV$) is varied by changing the distance
between the two beams horizontally by $D_x$ and vertically by $D_y$
($x-y$ being the plane transverse to the beam axis).
The $x$ and $y$ plane scans are performed individually.
Then the functional shape $\RV(D_x,0)$ and $\RV(0, D_y)$, and based on that
the transverse areas of the scan shapes are obtained as:
\begin{equation}
  S_x =  \int_{-\infty}^\infty \RV(D_x,0) dD_x \;\;\; \mbox{and} \;\;\;  S_y =  \int_{-\infty}^\infty \RV(0, D_y) dD_y.
\end{equation}
Analytical calculations \cite{VDMOYAMA} show that $S_{x,y}$ and $\RV$ at head-on have relations:
\begin{equation}
 \RV(0,0)/S_x = Q_x \;\;\; \mbox{and} \;\;\; \RV(0,0)/S_y = Q_y
\end{equation}
where the shape factor $Q_{x,y}$ 
correspond to $1/\sqrt{4\pi\sigma_{x,y}^2}$ if the beams have gaussian shape with sizes $\sigma_{x,y}$.
The $\sigma_{x,y}$ are obtained by a fitting method using gaussian function,
and compared to another method which calculates $S_{x,y}$ numerically without fitting.
With intensities of beams $N_1$ and $N_2$, the $L$ and hence $\SV$ are obtained by:
\begin{equation}
  L=k_b f N_1 N_2 Q_x Q_y \;\;\; \mbox{and} \;\;\; \SV= \RV(0,0) / L  \label{Eq:LumiandCS}
\end{equation}
where $k_b$ is the number of colliding bunches, and $f\sim11$~kHz is the orbital frequency.

During the vdM scan, 
the $N_{1,2}$ are monitored by LHC beam instruments based on inductive current pickup devices.
The calibration and correction results with systematic uncertainties are provided by a working group
formed by LHC and experiments \cite{BCNWG-CERNREF}.

\begin{figure*}[tb]
  \centering
  \includegraphics*[width=9.5cm]{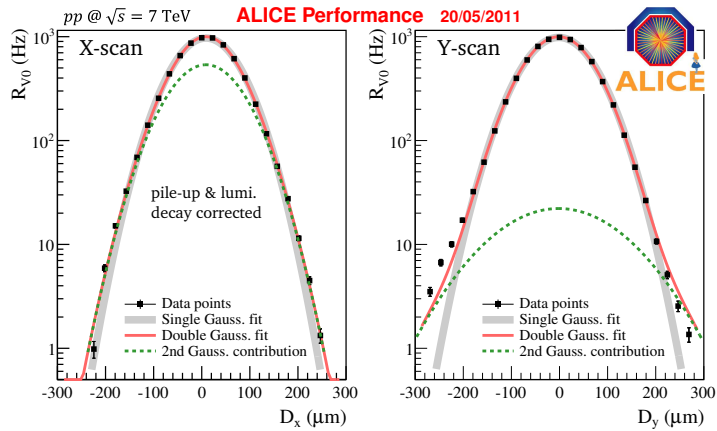}
  \caption{Corrected data for X-scan (left) and Y-scan(right).
           Fit results by single gaussian (bold line) and double gaussian (solid line) are overlaid.
           The dashed lines show the secondary component of the double gaussian fit function.
           \label{Fig:FIT-BC}
  }
\end{figure*}

Bunch crossings in which more than one interaction occurs are still counted as one trigger (pile-up effect).
The average number of interactions per bunch crossing is indicated by $\mu$.
Assuming Poisson statistics, $\RV$ is reduced by factor $( 1-e^{-\mu} )/\mu$.
This correction (up to 40\%) has been applied to the individual rate data\cite{VDMOYAMA}.

The calibration of the separation scale has been verified by taking data when
both beams were moved with identical displacement.
The small movement of the event vertex distribution was measured by the experiments and used for correction (up to 2.2\%).

The results after those corrections, with few more minor corrections, are shown in Fig.\ref{Fig:FIT-BC}.
While the single gaussian fit is acceptable for the $x$-scan,
non-gaussian and asymmetric tails are observed for the $y$-scan.

\section{Results}

\begin{figure}[tb]
  \centering
  \includegraphics*[width=15cm]{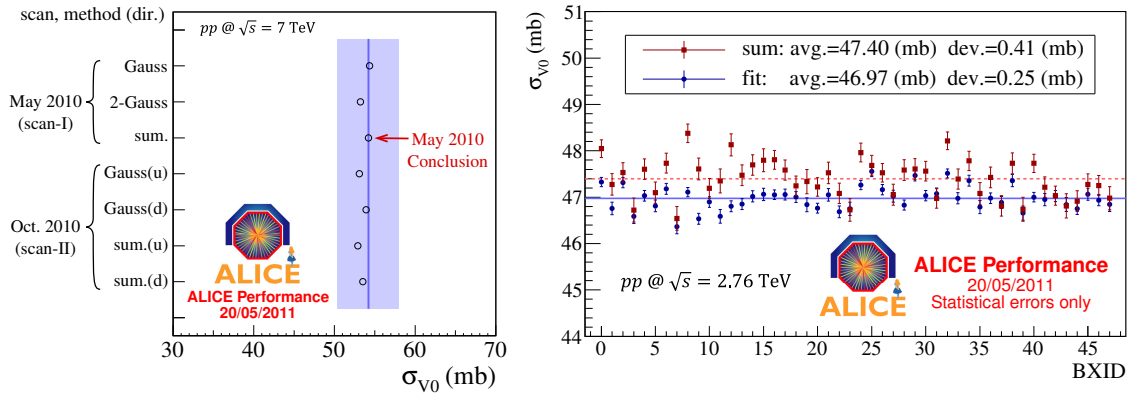}
  \caption{
    Left: comparison between two 7 TeV scans performed in May and Oct. 2010, and among different methods;
    gauss and double gauss fits, and scan directions (`u' for up, and `d' for down).
    The final value extracted from the May result and its systematic uncertainty are indicated
    by the vertical line and the gray band.
    Right: measured cross section for all colliding bunch pairs in at 2.76 TeV.
    Numerical sum and gaussian fit methods are compared.
    \label{Fig:CS_RESULT}
  }
\end{figure}

Fig.~\ref{Fig:CS_RESULT} shows results of three vdM scans performed in ALICE.
For 7 TeV, two scans in May (finalized) and Oct. (preliminary) in 2010 are compared.
To take the tails in $\RV$ distributions into account,
the result by the sum method (54.2~mb) was accepted as the final value\cite{VDMOYAMA}.
For 2.76 TeV, one scan was performed in Mar. 2011 with 48 bunch pair collisions in ALICE.
Although $N_1N_2$ varies up to factor 2, the pair-to-pair deviation is below 0.5\%, indicating the pile-up correction works.
The average between the sum and the fitting method (47.2~mb) was accepted as preliminary result for 2.76 TeV.

The largest source of systematic uncertainties are beam current uncertainties.
For 7 TeV and 2.76 TeV, it is 4.4\%, and 5\%, respectively for $\SV$ \cite{BCNWG-CERNREF}.
The second largest source of uncertainties is the distance scale uncertainty which is 2$\%\oplus$2\% for $\SV$.
Together with other smaller uncertainties, total uncertainties are estimated to be 7\% for both energies.

The obtained $\SV$ was used to determine cross sections
of several processes such as $J/\psi$ production cross section\cite{JPSI}, inelastic cross section,
and etc (see other contributions of this proceedings).
As an important check, the cross section of processes with at least one primary charged particle in the kinematic volume of
$p_t>0.5$GeV/$c$ and $-0.8<\eta<0.8$ has been measured to be 42.4$\pm$2.0~mb.
The same measurements were performed in other experiments of LHC; ATLAS and CMS.
Those results are are 42.3$\pm$0.7~mb, 43.99$\pm$0.62~mb, respectively and are compatible to ALICE result.

\section{Conclusion}\label{Sec:Conclusion}

The reference cross section ($\SV$) of the ALICE V0 detector
has been measured for 7 TeV and 2.76 TeV $pp$ collisions to be
$\SV(\mbox{7 TeV})    =  54.2 \pm 3.8\mbox{(syst.)} \; \mbox{mb}$ and
$\SV(\mbox{2.76 TeV}) =  47.2 \pm 3.3\mbox{(syst.)} \; \mbox{mb}$ (preliminary).
The results were used for cross section determinations of other physics processes in ALICE.


\section*{References}
\begin{thebibliography}{10}

\bibitem{ALICE}
  The ALICE Collaboration and K Aamodt et al., 2008 {\it JINST} {\bf 3} S08002

\bibitem{VDM}
  S.~van~der~Meer, ISR-PO/68-31, KEK68-64

\bibitem{VDMOYAMA}
  K.~Oyama, CERN-Proceedings-2011-001, p39

\bibitem{BCNWG-CERNREF}
  G.~Anders et al., CERN-ATS-Note-2011-004 PERF, 31 Jan 2011

\bibitem{JPSI}
  The ALICE Collaboration and K. Aamodt et al., 2011 arXiv:1105.0380v1 [hep-ex]

\end{thebibliography}
\end{document}